\documentclass[onecolumn,superscriptaddress,showpacs,preprintnumbers,amssymb]{revtex4}
\usepackage{amsmath}
\usepackage{graphicx}
\usepackage{dcolumn}
\usepackage{bm}
\def\jpsi{{J/\psi}}

\def\be{\begin{equation}}
\def\ee{\end{equation}}
\def\bea{\begin{eqnarray}}
\def\eea{\end{eqnarray}}
\makeatletter
\def\hlinew#1{%
  \noalign{\ifnum0=`}\fi\hrule \@height #1 \futurelet
   \reserved@a\@xhline}
\def\NO{\nonumber}
\def\gev{\mathrm{~GeV}}
\def\fb{\mathrm{~fb}}
\def\dfrac{\displaystyle\frac}

\def\co{{\cal O}}
\def\a{\alpha}
\def\b{\beta}

\def\e{\epsilon}
\def\g{\gamma}
\def\s{\sigma}

\begin{document}


\title{ ${\mathcal O}(\alpha_s v^2)$
correction to $J/\psi$ plus $\eta_c$ production in $e^+e^-$ annihilation
at $\sqrt{s}=10.6GeV$}

\author{Xi-Huai Li}
\affiliation{
Department of Modern Physics, University of Science and Technology of China, Hefei, Anhui, 230026,China.}
\affiliation{Institute of High Energy Physics, Chinese Academy of
Science, P.O. Box 918(4), Beijing, 100049, China.}

\author{Jian-Xiong Wang}
\affiliation{Institute of High Energy Physics, Chinese Academy of
Science, P.O. Box 918(4), Beijing, 100049, China.}

\date{\today}

\begin{abstract}
Based on the nonrelativistic QCD factorization approach,
${\mathcal O}(\alpha_s v^2)$ corrections to $\jpsi$ plus $\eta_c$ production in $e^+e^-$ annihilation at $\sqrt{s}=10.6 \gev$
is calculated in this work. The numerical results show that the correction at $\alpha_s v^2$ order is only about a few percent for the
total theoretical result. It indicates that the perturbative expansions for the theoretical prediction become convergence and
higher order correction will be smaller. The uncertainties from the long-distance matrix elements, renormalization scale and
the measurement in experiment
are also discussed. Our result is in agreement with previous result in ref~\cite{Dong:2012xx}.
\end{abstract}

\pacs{12.38.Bx, 13.25.Gv, 13.60.Le}
\maketitle
\section{Introduction}
Study on heavy quarkonium decay and production is a very important and interesting issue to understand
quantum chromodynamics (QCD), the fundamental theory of strong interactions. Many experimental and theoretical reserches
have been performed since  the discovery of the $J/\psi$ charmonium meson in 1974 followed by the $\Upsilon$ bottomonium
meson in 1977, for reviews see Ref.\cite{Brambilla:2010cs}. In experimental side, it is easy to detect $J/\psi$ and $\Upsilon$
signal. In theoretical side, quarkonium bound states offer a solid ground to probe QCD, due to the
high scale provided by the large mass of the heavy quarks, which make the QCD factorization possible in the related calculation.
To explain the large discrepancy on the transverse momentum distribution of chromonium hadroproduction between the experimental
measurement and theoretical prediction as well as to arrange the infrared divergence cancellation in p-wave quarkonium related
calculation,   the nonrelativistic QCD (NRQCD) factorization approach\cite{Bodwin:1994jh} has been
introduced. It allows consistent theoretical prediction to be made and to be improved perturbatively in the QCD coupling
constant $\a_s$ and the heavy-quark relative velocity $v$ in heavy quarkonium rest frame.

In last five years, most of the important theoretical studies on heavy quarkonium based on NRQCD are calculated at next-to-leading
order (NLO) of QCD and many of them are also calculated at next-to-leading order of $v$. Among them, the $J/\psi$ polarization puzzle
at hadron colliders is still unclear after the important progresses at QCD NLO~\cite{Campbell:2007ws},
It seems that the inclusive $J/\psi$ production
at B-factories can be explained by just color singlet contribution at QCD 
NLO~\cite{He:2007te,Zhang:2006ay},
but it causes the problem for
the color-octet long distance matrix elements~\cite{Zhang:2009ym}. The theoretical calculation with NLO QCD and relativistic correction
can cover the experimental measurements on exclusive double chromonium production at B-factories although the corrections are very large.
For theoretical prediction based on perturbative expansion, the convergence of the expansion is a very important issue.
Therefore, it is important to test the calculation at higher order when the NLO correction is large. Usually, higher order calculation
is much more complicate,  so far there are only a few simple processes whose $\mathcal O(\alpha_s v^2)$ corrections are
calculated \cite{Luke:1997ys,Guo:2011tz,Jia:2011ah}.

For the exclusive double chromonium production at B-factories, its higher order calculation is studied, so we give a detailed review on it.
The exclusive production cross section of double charmonium in $e^+e^-\rightarrow \jpsi\eta_c$ at $\sqrt{s}=10.6$ GeV
measured by Belle \cite{Abe:2002rb,Pakhlov:2004au} is
$\s[\jpsi+\eta_c] \times B^{\eta_c}[\geq2] = (25.6\pm2.8\pm3.4)\fb$
and by BABAR \cite{Aubert:2005tj} is
$\s[\jpsi+\eta_c] \times B^{\eta_c}[\geq2] = (17.6\pm2.8^{+1.5}_{-2.1})\fb$,
where $B^{\eta_c}[\geq2]$ denotes the branching fraction for the $\eta_c$ decaying into at least two charged tracks.
Meanwhile, the NRQCD LO theoretical predictions in the QCD coupling constant $\a_s$ and the charm-quark relative velocity $v$,
given by Braaten and Lee \cite{Braaten:2002fi}, Liu, He and Chao \cite{Liu:2002wq}, and  Hagiwara, Kou and Qiao \cite{Hagiwara:2003cw}
are about $2.3 \sim 5.5 \fb$,
which is an order of magnitude smaller than the experimental results. Such a large discrepancy between experimental results
and theoretical predictions brings a challenge to the current understanding of charmonium production based on NRQCD.
Many studies have been performed in order to resolve the problem. From treatments beyond NRQCD,
Ma and Si \cite{Ma:2004qf} treated the process by using light-cone method,
a similar treatment was performed by Bondar and Chernyad \cite{bondar:2005} and Bodwin, Kang and Lee \cite{Bodwin:2006dm},
possible contribution from intermediate meson rescatterings was  considered by Zhang, Zhao, and Qiao \cite{Zhang:2008ab},
it was also studied in the Bethe-Salpeter formalism by Guo, Ke, Li, and Wu in Ref~\cite{Guo:2008cf}.
Based on NRQCD,
Braaten and Lee \cite{Braaten:2002fi} have shown that the relativistic corrections would increase the cross section by
a factor of about 2,
and the NLO QCD correction of the process has been studied by Zhang, Gao and Chao \cite{Zhang:2005ch}
and Gong and Wang \cite{Gong:2007db}, which
can enhance the cross section with a $K$ factor (the ratio of NLO to LO) of about 2,
again the relativistic corrections have been studied by Bodwin, Kang, Kim, Lee and Yu \cite{Bodwin:2006ke}
and by He, Fan and Chao \cite{He:2007te}, which is significant.
More detailed treatment, such as including the resummation of a class of relativistic correction, has been taken into
consideration by Bodwin and Lee and Yu \cite{Bodwin:2007ga}. 
In another way, Bodwin, Lee and Braaten \cite{Bodwin:2002fk} showed that the cross section for the process 
$e^+e^- \rightarrow \jpsi + \jpsi$ may be larger than that for $\jpsi + \eta_c$ by a factor of 1.8, in spite 
of a suppression factor $\a^2/\a_s^2$ that is associated with the QED and QCD coupling constants. 
They suggested that a significant part of the discrepancy of $\jpsi+\eta_c$ production may be explained by this process.
Hagiwara, Kou and Qiao \cite{Hagiwara:2003cw} also calculated and discussed this process. In 2004, a new analysis of double 
charmonium production in $e^+e^-$ annihilation was performed by Belle \cite{Abe:2004ww} based on a 3 times larger data set 
and no evidence for the process $e^+e^- \rightarrow \jpsi + \jpsi$ was
found. Both the NLO QCD corrections and relativistic corrections to $e^+e^-\rightarrow\jpsi+\eta_c$ give a large K factor of about 2. It is
obvious that these two types of corrections to $e^+e^-\rightarrow\jpsi+\jpsi$ should be studied to explain the experimental results. 
In fact, they have been studied by Bodwin, Lee and Braaten for the dominant photon-fragmentation contribution diagrams 
\cite{Bodwin:2002kk}. The results show that the cross section is decreased by K factor of 0.39 and 0.78 for the NLO QCD and 
relativistic corrections respectively.
A more reliable estimate, $1.69\pm 0.35$ fb, was given by Bodwin, Lee, Braaten and Yu in ref.~\cite{Bodwin:2006yd}.
And light-cone method is used in ref.~\cite{Braguta:2007ge} by V.V. Braguta.
Gong and Wang performed a complete NLO QCD calculation on $e^+e^- \rightarrow \jpsi + \jpsi$~\cite{Gong:2008ce} and the results show that 
the cross section would be much smaller than the rough estimate in Ref.~\cite{Bodwin:2002kk}.
Therefore it is easy to understand why there was no evidence for the process $e^+e^- \rightarrow \jpsi + \jpsi$ at B-factories.

It is easy to see that
both the QCD correction ($\alpha_s$) and relativistic correction ($v^2$)  are very large for
$e^+e^-\rightarrow \jpsi\eta_c$ at B-factory energy,
and with these corrections the experimental measurement can be explained.
Therefore it is natural to ask the question, how is the situation for the higher order corrections beyond $\alpha_s$ and $v^2$ correction
?  $\alpha_s^2$ correction is very difficult to do, but recent progress make it available to do $\alpha_s v^2$ correction already.
It is very interesting to see that the $\alpha_s v^2$ correction, given in a recent work~\cite{Dong:2012xx}, is a small contribution.
It convinces us in some sense (with $\alpha_s^2$ correction absent) that the double expansions in NRQCD converges quite well
on this problem.
Since the calculation is quite complicate and plays an important role to convince us the convergence on the theoretical predication
which can explain the experimental data, in this paper we performed an independent calculation on it by using the
our package Feynman Diagram Calculation (FDC)~\cite{Wang:2004du} with the built-in method to calculate relativistic correction.
The remainder of this paper is organized
as follows. Base on the NRQCD frame, we briefly introduce theoretical formulism for the
calculation of heavy quarkonium production and give the corresponding results in perturbative NRQCD in Sec.~\ref{theory}. 
The details in perturbative QCD are summarized in Sec.~\ref{qcd}. We give the numerical results of $\alpha_s v^2$ corrections 
and some discussion in Sec.\ref{results}.
Finally, in Sec.~\ref{summary}, we present a brief summary.

\section{NRQCD FACTORIZATION FORMULA up to $v^{2}$ order}
\label{theory}
According to  NRQCD effective theory, the production of the charmonium are factorized into
two parts, the short-distance part and the long-distance part. The long-distance parts are related to the
four fermion operators, characterized by the velocity $v$ of the
charm quark  in the meson rest frame. The long-distance matrix elements  can be estimated by lattice calculations or phenomenological
models, or determined by fitting experimental data. The production cross section
up to $v^{2}$ order is expressed as

\begin{equation}
\sigma(e^+e^-\to J/\psi+\eta_c)=(c_{00}+c_{10}\langle v^2\rangle_{J/\psi}+c_{01}\langle v^2\rangle_{\eta_c})\langle{\mathcal{O}_1}\rangle_{\eta_c}
\langle{\mathcal{O}_1}\rangle_{J/\psi}
\end{equation}
with the long-distance matrix elements being defined by using related operators as
\bea
\langle v^2\rangle_{J/\psi} =\dfrac{\langle{\mathcal{P}_1}\rangle_{J/\psi}}{m_c^2\langle{\mathcal{O}_1}\rangle_{J/\psi}},~ ~ ~ ~ ~ ~
\langle{\mathcal{O}_1}\rangle_{J/\psi}=\langle 0|\chi^{\dagger}\sigma^{i}
\psi(a_{J/\psi}^{\dagger}a_{J/\psi})\psi^{\dagger}\sigma^{i}\chi|0\rangle, \\ \NO
\langle{\mathcal{P}_1}\rangle_{J/\psi}=\langle 0|\frac{1}{2}[\chi^{\dagger}\sigma^{i}\psi(a_{J/\psi}^{\dagger}a_{J/\psi})
\psi^{\dagger}\sigma^{i}(-\frac{i}{2}\overleftrightarrow{\mathbf{D}})^2\chi+
\chi^{\dagger}\sigma^{i}(-\frac{i}{2}\overleftrightarrow{\mathbf{D}})^2\psi(a_{J/\psi}^{\dagger}a_{J/\psi})
\psi^{\dagger}\sigma^{i}\chi]|0\rangle.
\eea
for $J/\psi$ and
\bea
\langle v^2\rangle_{\eta_c}=\dfrac{\langle{\mathcal{P}_1}\rangle_{\eta_c}}{m_c^2\langle{\mathcal{O}_1}\rangle_{\eta_c}},~ ~ ~ ~ ~ ~
\langle{\mathcal{O}_1}\rangle_{\eta_c}=\langle 0|\chi^{\dagger}\psi(a_{\eta_{c}}^{\dagger}a_{\eta_{c}})\psi^{\dagger}\chi|0\rangle, \\ \NO
\langle{\mathcal{P}_1}\rangle_{\eta_c}=\langle 0|\frac{1}{2}[\chi^{\dagger}\psi(a_{\eta_{c}}^{\dagger}a_{\eta_{c}})
\psi^{\dagger}(-\frac{i}{2}\overleftrightarrow{\mathbf{D}})^2\chi+
\chi^{\dagger}(-\frac{i}{2}\overleftrightarrow{\mathbf{D}})^2\psi(a_{\eta_{c}}^{\dagger}a_{\eta_{c}})
\psi^{\dagger}\chi]|0\rangle,
\eea
for $\eta_c$ where $m_c$ is the charm quark mass.
It is the basic point that the NRQCD factorization for hadron related process will also hold when
the hadron state are replaced by $Q\bar{Q}$ states  with exactly the same quantum numbers as the corresponding hadron state.
In this way, the short-distance coefficients $c_{00},c_{01}$ and $c_{10}$ can be obtained in perturbative calculation through
the matching condition, and they are calculated up to  QCD next-to-leading (NLO) order.
In order to obtain the short-distant coefficients, the matrix elements of the operators for quantum states need to be calculated
perturbatively, and there are
\begin{eqnarray}
\langle{\mathcal{O}_1}\rangle_{{}^{1}S_{0}} =2N_c(2E_{q_1})^2,~ ~ ~ ~
\langle{\mathcal{O}_1}\rangle_{{}^{3}S_{1}} =6N_c(2E_{q_2})^2
\end{eqnarray}
where there are $N_c=3$ for $SU(3)$ group and $E_q=\sqrt{m_c^2+q^2}$.
From the NRQCD effective Lagrangian, we could easily get the Feynman rules. Therefore
we  have calculated order $\alpha_s v^2$ corrections to the leading order $\langle{\mathcal{O}_1}\rangle_{^{2s+1}S_{s}}$ in
perturbative NRQCD with the dimensional regularization and
defined the renormalization constants $Z_O^{\overline{\mathrm{MS}}}$ of the operator by using the
$\overline{\textrm{MS}}$ scheme \cite{Bodwin:1994jh,Bodwin:1998mn}.
\be
\delta Z_O^{\overline{\mathrm{MS}}}~~=~~-\frac{4\alpha_s C_F}{3\pi}(\frac{\mu_r^2}{\mu_{\Lambda}^2})^{\epsilon}(\frac{1}{\epsilon_{\textrm{UV}}}+\ln{4\pi}-\gamma_\textrm{E})
\frac{\mathbf{q}^2}{m_c^2}
\ee

\begin{eqnarray}
\langle{\mathcal{O}_1}\rangle_{{}^{2s+1}S_{s}}^{\textrm{R}}
  = [1+
\frac{4\alpha_s C_F}{3\pi}(\frac{
\mu_r^2}{\mu_{\Lambda}^2})^{\epsilon}
(\frac{1}{\epsilon}+\ln{4\pi}-\gamma_\textrm{E})
\frac{\mathbf{q}^2}{m_c^2}]\langle{\mathcal{O}_1}\rangle_{^{2s+1}S_{s}}.
\end{eqnarray}

\begin{eqnarray}
\langle{\mathcal{P}_1}\rangle_{{}^{2s+1}S_{s}}=
\mathbf{q}^2 \langle{\mathcal{O}_1}\rangle_{^{2s+1}S_{s}}.
\end{eqnarray}
At last we could easily give the perturbative NRQCD results.
\begin{eqnarray}\label{NRQCD}
\sigma(e^+e^-\to Q\bar{Q}(^{3}S_1^1)+Q\bar{Q}(^{1}S_0^1))\Big{|}_{\textrm{pertNRQCD}}\!&=&\!
 \{c_{00}+\frac{\mathbf{q_1}^2}{m_c^2}[c_{10}+
\frac{4\alpha_s C_F}{3\pi}(\frac{
\mu_r^2}{\mu_{\Lambda}^2})^{\epsilon}(\frac{1}{\epsilon}+\ln{4\pi}-\gamma_\textrm{E})c_{00}^0
] +\frac{\mathbf{q_2}^2}{m_c^2}\nonumber\\
&&\![c_{01}+\frac{4\alpha_s C_F}{3\pi}(\frac{
\mu_r^2}{\mu_{\Lambda}^2})^{\epsilon}(\frac{1}{\epsilon}+\ln{4\pi}-\gamma_\textrm{E})c_{00}^0
] \}{192(N_cE_{q_1}E_{q_2})^2}
\end{eqnarray}

\section{details of perturbative QCD calculation} \label{qcd}

For a $Q(p)\bar{Q}(\bar{p})$ quantum state, we denote $P$ as the total momentum and $q$ as the relative momentum between
$Q$ and $\bar{Q}$ pair. Therefore, there are
\bea
p=\frac{1}{2}P+q, ~ ~ ~ ~
\bar{p}=\frac{1}{2}P-q. \\ \NO
p^2=\bar{p}^{\,2}=m_Q^2,~ ~ ~ ~
P^2=4E_q^2,~ ~ ~E_q=\sqrt{m_Q^2+\bm{q}^2}
\eea
where $m_Q$ is the mass of the heavy quark $Q$, and the $Q$
and $\bar{Q}$ are on their mass shells.

To do the perturbative calculation in related process for the quantum states, we could obtain
the projectors for each quantum states.
The spin-singlet and spin-triplet components of each $Q\bar{Q}$ state can be
projected out by making use of the spin projectors. After multiplying
corresponding Clebsch-Gordan coefficients to the spin component of the
outer product of the spinors for each $Q\bar{Q}$ pair, one can find that
 $\bar{\Pi}_1$ and $\bar{\Pi}_3$ are the
spin-singlet and spin-triplet projectors of the $Q\bar{Q}$ production,
respectively. The spin projectors that are valid to all orders in the relative
momentum can be found in Refs\cite{Bodwin:2002hg}.

\begin{subequations}
\label{spin-pro-1}
\begin{eqnarray}
\Pi_1&=&\phantom{-}\frac{1}{4\sqrt{2}E(E+m_Q)}
(\,/\!\!\!\bar{p}-m_Q)\,\gamma_5(\,/\!\!\!\!P+\!2E)\,(\,/\!\!\!{p}+m_Q),\\
\Pi_{3}&=&\phantom{-}\frac{1}{4\sqrt{2}E(E+m_Q)}
(\,/\!\!\!\bar{p}-m_Q)\,/\!\!\!\epsilon^*(\lambda)(\,/\!\!\!\!P+\!2E)\,
(\,/\!\!\!{p}+m_Q),
\end{eqnarray}
\end{subequations}
where $\Pi_1$ and $\Pi_{3}$  are projectors for spin 0 and spin 1 s-wave quantum states respectively,
and $\epsilon^*(\lambda)$ is the polarization vector of the spin-triplet state.

For process $e^+(p_1)e^-(p_2)\to Q(\frac{p_3}{2}-q_1)\bar{Q}(\frac{p_3}{2}+q_1)(^{3}S_1^1)
+Q(\frac{p_4}{2}-q_2)\bar{Q}(\frac{p_4}{2}-q_2)(^{1}S_0^1)$, the production matrix element is expressed as
\begin{eqnarray}
&&{\cal M}\left(e^+e^-\to Q\bar{Q}(^{3}S_1^1)
+Q\bar{Q}(^{1}S_0^1)\right)=\epsilon_{\mu}(S_z) A^{\mu}(q_1,q_2) \nonumber\\
&&=\epsilon_{\mu}(S_z)\left( A^{\mu}\Big{|}_{q_1=0,q_2=0}+\frac{q_1^2}{6}I^{\alpha\beta}\frac{\rm d^{3}A^{\mu}}{\rm d
q_1^{\alpha}\rm d q_1^{\beta}}\Big{|}_{q_1=0,q_2=0}+\frac{q_2^2}{6}I^{\alpha\beta}\frac{\rm d^{3}A^{\mu}}{\rm d
q_2^{\alpha}\rm d q_2^{\beta}}\Big{|}_{q_1=0,q_2=0} \right) + \co(q_1^4,q_2^4)
\end{eqnarray}
where we have used the following relation
\be
\label{amp-spin-swave-v4}
\int\!\dfrac{d\Omega}{4\pi}q^\mu =0, ~ ~ ~
\int\!\dfrac{d\Omega}{4\pi}q^\mu q^\nu =\frac{\bm{q}^2}{3}I^{\mu\nu}, ~ ~ ~
{\alpha\beta}=-g^{\alpha\beta}+\frac{P^{\alpha}P^{\beta}}{P^2}.
\ee
As for the expansion of $q$, we should consider the effect that the external momentum
and polarization vector may be the implicit function of $q$. 
From the momentum conservation and on-shell conditions, $p_3^2=4E_{q_1}^2,p_4^2=4E_{q_2}^2$,
we could find that $p_3,~p_4$ are implicit functions of $q_1,~q_2$ respectively.
However,it is obvious that the short-distance coefficients, to be obtained in the perturbative calculation,
are functions of the independent variables which are the invariant mass $s$ of the $e^+$ and $e^-$ system and
$\cos\theta$. $\theta$ is the angle between $J/\psi$ and the electron.
Where $s$ and $cos\theta$ are independent of the relative momentum $q$. 

Since the final results are  Lorentz invariance and irrelevant to
the reference  frame, we choose to do the calculation in the center-of-mass of this system
where $p_1+p_2=p_3+p_4=(\sqrt{s},0,0,0)$ is the explicit expression of the momentum conservation.
Therefore the following results are obtained:

\be
\label{dpdq}
\dfrac{\rm d p_3}{\rm d q^2_1}\cdot p_3=2,~ ~ ~ ~\dfrac{\rm d p_4}{\rm d q^2_1}\cdot p_4=0,~ ~ ~ ~ 
\dfrac{\rm d p_3}{\rm d q^2_1}+\dfrac{\rm d p_4}{\rm d q^2_1}=0,~ ~ ~ ~\dfrac{\rm d p_3}{\rm d q^2_1}\cdot p_4=0.
\ee
We choose two vectors $r_1=(0,\overrightarrow{r_1})$ and $r_2=(0,\overrightarrow{r_2})$ with
$\overrightarrow{r_1}$ and $\overrightarrow{r_2}$ being unit vectors, while
$\overrightarrow{r_1}, \overrightarrow{r_2}$ and $\overrightarrow{p_3}$ are perpendicular to each other, i.e
$r_1\cdot r_2=0,p_3\cdot r_1=0,p_3\cdot r_2=0$. Then vector $\dfrac{\rm d p_3}{\rm d q^2_1}$ can be expressed as linear combination of
four independent vectors as $\dfrac{\rm d p_3}{\rm d q^2_1}=a_1 p_3 + a_2 p_4 + a_3 r_1 +a_4 r_2$. From the following
conditions
\be
\dfrac{\rm d p_3}{\rm d q^2_1}\cdot r_1=0,~ ~ ~\dfrac{\rm d p_3}{\rm d q^2_1}\cdot r_2=0
\ee
together with previous conditions in Eq.(\ref{dpdq}), we can easily obtain the solution
\be
\dfrac{\rm d p_3}{\rm d q^2_1}=\dfrac{-2p_4^2}{(p_3\cdot p_4)^2-p_3^2p_4^2}p_3+\dfrac{2p_3\cdot p_4}{(p_3\cdot p_4)^2-p_3^2p_4^2}p_4.
\ee
For the $\epsilon^*(\lambda)$, the polarization four-vector of the $|Q\bar{Q}({}^{3}S_{1})\rangle$ with helicity $\lambda$,
there are the relation $\dfrac{\rm d\epsilon^*(\pm1)}{\rm d q^2_1}=0$ since $\theta$ is independent of the relative momentum $q$.
It is easy to obtain
\be
\dfrac{\rm d \epsilon^*(0)}{\rm d q^2_1}\cdot p_3=-\dfrac{\rm d p_3}{\rm d q^2_1}\cdot \epsilon^*(0),~ ~
\dfrac{\rm d \epsilon^*(0)}{\rm d q^2_1}\cdot \epsilon^*(0)=0,~ ~
\dfrac{\rm d \epsilon^*(0)}{\rm d q^2_1}\cdot \epsilon^*(1)=0,~ ~
\dfrac{\rm d \epsilon^*(0)}{\rm d q^2_1}\cdot \epsilon^*(-1)=0.
\ee
Therefore, we obtain the relation between the polarization four-vector and $q$ as
\be
\dfrac{\rm d \epsilon^*(\lambda)}{\rm d q^2_1}=\dfrac{-\dfrac{\rm d p_3}{\rm d q^2_1} \cdot \epsilon^*(\lambda)p_3}{p_3^2}=
\dfrac{-2 p_3\cdot p_4 p_4\cdot \epsilon^*(\lambda)p_3}{((p_3\cdot p_4)^2-p_3^2p_4^2)p_3^2}.
\ee
The treatment about $q_2$ is similar to these.

Considering the two body phase space, we need to expand it.
$$d\Gamma=\int\!d\cos\!\theta\dfrac{2|\overrightarrow{p'}|}{16\pi\sqrt{s}}$$
where $|\overrightarrow{p'}|=\dfrac{\lambda^{1/2}(s,4E_{q_1}^2,4E_{q_2}^2)}{2\sqrt{s}},\lambda(x,y,z)=x^2+y^2+z^2-2(xy + yz + xz)$.
Only $|\overrightarrow{p'}|$ need to be expanded since $\cos\theta$ and $s$ are independent of $q_1$ and $q_2$. Then there is
$$d\Gamma=\int\!d\cos\!\theta\dfrac{2|\overrightarrow{p}|}{16\pi\sqrt{s}}(1-\dfrac{1}{|\overrightarrow{p}|^2}(q_1^2+q_2^2))$$
 where $|\overrightarrow{p}|=\dfrac{\lambda^{1/2}(s,4m^2,4m^2)}{2\sqrt{s}}$. We could square the amplitude,
 integrate over the phase space, and expand in powers of $q$ in order to obtain the desired perturbative result

\be
\sigma(e^+e^-\to Q\bar{Q}({}^{3}S_{1}^1)+Q\bar{Q}({}^{1}S_{0}^1))\Big{|}_{\textrm{pertQCD}}=\int\!d\Gamma\sum_{s_z}
{\Big{|}\cal M\Big{|}}^2.
\ee

Most of the steps in this section are realized in a small program in FDC package, and the final Fortran source for numerical calculation
are prepared by using FDC package together with the small program for $q^2$ expansion.

Since there is no $\co(\alpha_s)$ real process in NLO, we only need to calculate virtual corrections.  Dimensional regularization has been
adopted for isolating the ultraviolet(UV) and infrared(IR) singularities.
UV divergences  are cancelled upon the renormalization of the QCD gauge coupling constant, the charm quark mass and field, and the gluon field. A similar renormalization scheme is chosen as in ref. \cite{Klasen:2004tz} except that both light quarks and charm quark are included in the quark loop to obtain the renormalization constants. The renormalization constants of the charm quark mass $Z_m$ and field $Z_2$, and the gluon field $Z_3$ are defined in the on-mass-shell(OS) scheme while that of the QCD gauge coupling $Z_g$ is defined in the modified-minimal-subtraction($\overline{\mathrm{MS}}$) scheme:
\bea
\delta Z_m^{OS}&=&-3C_F\dfrac{\alpha_s}{4\pi}\left[\dfrac{1}{\e_{UV}} -\gamma_E +\ln\dfrac{4\pi \mu^2}{m_c^2} +\frac{4}{3} +\co(\e)\right] , \NO\\
\delta Z_2^{OS}&=&-C_F\dfrac{\alpha_s}{4\pi}\times\biggl[\dfrac{1}{\e_{UV}} +\dfrac{2}{\e_{IR}} -3\gamma_E +3\ln\dfrac{4\pi \mu^2}{m_c^2} +4 +\co(\e)\biggr] , \NO\\
\delta Z_3^{OS}&=&\dfrac{\alpha_s}{4\pi}\biggl[(\beta'_0-2C_A)\left(\dfrac{1}{\e_{UV}} -\dfrac{1}{\e_{IR}}\right)
-\dfrac{4}{3}T_F\left(\dfrac{1}{\e_{UV}} -\gamma_E +\ln\dfrac{4\pi \mu^2}{m_c^2}\right) +\co(\e)\biggr] , \NO\\
\delta Z_g^{\overline{\mathrm{MS}}}&=&-\dfrac{\beta_0}{2}\dfrac{\alpha_s}{4\pi}\left[\dfrac{1}{\e_{UV}} -\gamma_E +\ln(4\pi) +\co(\e)\right] .
\eea
where $\g_E$ is Euler's constant, $\b_0=\frac{11}{3}C_A-\frac{4}{3}T_Fn_f$ is the one-loop coefficient of the QCD beta function and $n_f$ is the number of active quark flavors. There are three massless light quarks $u, d, s$, and one heavy quark $c$, so $n_f$=4. In $SU(3)_c$, color factors are given by $T_F=\frac{1}{2}, C_F=\frac{4}{3}, C_A=3$. And $\b'_0\equiv\b_0+(4/3)T_F=(11/3)C_A-(4/3)T_Fn_{lf}$ where $n_{lf}\equiv n_f-1=3$ is the number of light quarks flavors. Actually in the NLO total amplitude level, the terms proportion to $\delta {Z_3}^{OS}$  cancel each other, thus the result is independent of renormalization scheme of the gluon field.

\section{Results}
\label{results}
The final results are obtained by using the matching method with the UV and  IR divergences being cancelled.
\bea
\sigma=\sigma_{LO}+\sigma_{NLO(\alpha_s)}+\sigma_{NLO(v^2)}+\sigma_{NLO(\alpha v^2)}
\eea
$\sigma_{LO},\sigma_{NLO(\alpha_s)},\sigma_{NLO(v^2)},\sigma_{NLO(\alpha v^2)}$ are the contributions from the leading order,
the next leading order in $\alpha_s$ , the next leading in  $v^2$ and the next leading in $\alpha v^2$. Then
the production rate up to $\mathcal O(\alpha_s v^2)$ order is expressed as
\bea
\sigma=&\frac{8192\pi^3 C_F^2 e_c^2 \alpha_s^2(u_r)\alpha^2 (1-4 r)^{3/2}}{9N_c^2 s^4}
\langle{\mathcal{O}_1}\rangle_{\eta_c}\langle{\mathcal{O}_1}\rangle_{J/\psi}
\{1 + v_{J/\psi}^2 f_1(r) + v_{\eta_c}^2 f_2(r) + \frac{\alpha_s(\mu_r)}{\pi}[\beta_0 \ln{\frac{\mu_r}{2m_c}}
+ f_3(r)] \NO \\
 &+ \frac{\alpha_s(\mu_r)}{\pi}  v_{J/\psi}^2 [\beta_0 \ln{\frac{\mu_r}{2m_c}} f_1(r) +\frac{32}{9}\ln{\frac{\mu_\Lambda}{m_c}} + f_4(r)]
+\frac{\alpha_s(\mu_r)}{\pi} v_{\eta_c}^2 [\beta_0 \ln{\frac{\mu_r}{2m_c}} f_2(r)+ \frac{32}{9}\ln{\frac{\mu_\Lambda}{m_c}} + f_5(r)] \}
\eea
where there are $e_c=\frac{4}{3}$, $r=\frac{4m_c^2}{s}$, $f_1(r)=\frac{9-74r+80r^2}{6(1-4r)}$,
$f_2(r)=\frac{11-82r+80r^2}{6(1-4r)}$ and $u_r$ is the renormalization scale. Therefore, the obtained
analytic expressions of the $v^2$ correction is in agreement with that in the paper ~\cite{Dong:2012xx}.
At the same time, the analytic expression of $f_3(r)$ in the results of the $\alpha_s$ correction is also in agreement with that
in the paper ~\cite{Gong:2007db}. Since the analytic expressions of $f_4(r)$ and $f_5(r)$ in the $\mathcal O(\alpha_s v^2)$
correction are so lengthy that we just give the numerical results for them. In the numerical calculation, there are
\bea
&f_1=0.97466,~ f_2=1.3080,~ f_3=12.358,~  f_4=3.8382,~f_5=3.2537, ~ ~ ~ for~ r=\frac{4 \times 1.4^2}{10.58^2}; \NO \\
&f_1=0.87465,~ f_2=1.2098,~ f_3=11.806,~  f_4=2.0543,~f_5=2.6668, ~ ~ ~ for~ r=\frac{4 \times 1.5^2}{10.58^2}. \NO
\eea
\begin{table}
\begin{tabular}
{|>{\centering}p{0.20\textwidth}|>{\centering}p{0.10\textwidth}|>{\centering}p{0.10\textwidth}
|>{\centering}p{0.10\textwidth}|>{\centering}p{0.10\textwidth}|>{\hfill}p{0.10\textwidth}<{\hfill\hfill}|}
\hlinew{1pt}
   $\alpha_s(\mu_r)$  &  $\sigma_{LO}$  & $\sigma_{NLO(\alpha_s)}$ & $\sigma_{NLO(v^2)}$& $\sigma_{NLO(\alpha v^2)}$&$\sigma$ \\
\hline
   $\alpha_s(\frac{\sqrt{s}}{2}) = 0.211$  &  $4.381$  & $5.196$  & $1.714$ & $0.731$& $12.273$\\
\hline
   $\alpha_s(2 m_c) = 0.267$  &  $7.0145$  & $7.367$  & $2.745$ & $0.245$&$17.372$ \\
\hlinew{1pt}
\end{tabular}
\caption{With the follow parameters: $\alpha(\sqrt{s}) = 1/130.9$, $\langle
{\mathcal{O}_1}\rangle_{J/\psi}=1.161{\rm GeV}^3,\langle {\mathcal{O}_1}\rangle_{\eta_c}=0.387{\rm GeV}^3 $, $m_c=1.4$ GeV, $\langle v^2\rangle_{J/\psi} = 0.223$ , $\langle v^2\rangle_{\eta_c}= 0.133$, We give the cross sections with different renormalization scale $\mu$. Their units are fb.}
\label{tab:1}
\end{table}
\begin{figure}[tbH]
\begin{center}
\includegraphics[width=0.4\textwidth]{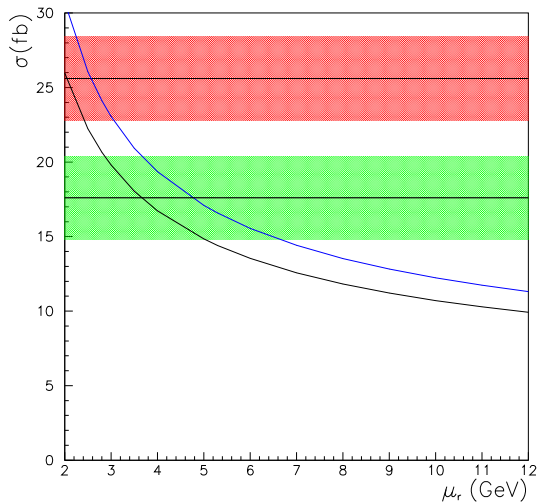}
\caption{The  cross section as a function of the  $\mu_r$ at $\sqrt{s}=10.58$ GeV. The black and blue solid curves are the
cross sections in the $m_c=1.4$ and $ m_c=1.5$  respectively. The red and green bands represent the measured
cross sections by the \textsc{Belle} and \textsc{BaBar} experiments,
with respective systematic and statistical errors. \label{figure1} }
\end{center}
\end{figure}
\begin{figure}[tbH]
\begin{center}
\includegraphics[width=0.4\textwidth]{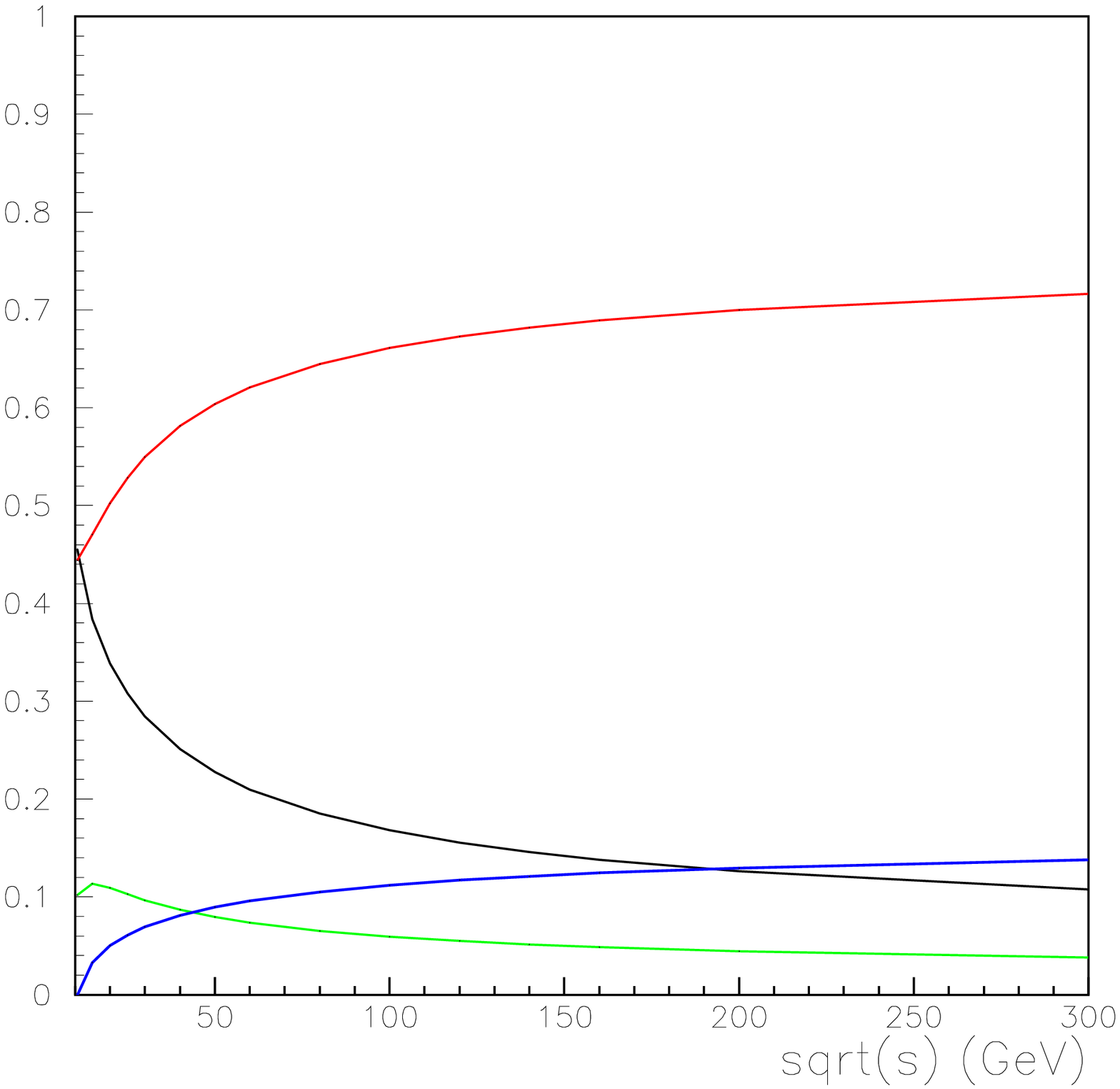}
\caption{The ratios of the different parts as a function of the $\sqrt{s}$. The black,red,green,blue lines are the ratios of the leading order,
$\alpha_s$ order, $v^2$ order,$\alpha_s v^2$ order respectively. \label{figure2}}
\end{center}
\end{figure}
And we take $\sqrt{s}=10.58$ GeV and $\mu_{\Lambda}=m_c$.
The running strong coupling constant is evaluated by using the two-loop formula with $\Lambda^{(4)}_{\overline{\rm MS}}= 0.338$
GeV as used in Ref~\cite{Gong:2007db}.
Our results are presented in the table~\ref{tab:1} with parameters given in table caption.
 The results are in agreement with that in ref~\cite{Dong:2012xx}. The contribution from the $\mathcal O(\alpha_s v^2)$ order is small.
 There are some  differences for the results in the table~\ref{tab:2} if the long-distance
matrices and QED coupling constant are chosen as in Ref~\cite{He:2007te}, the correction at $\mathcal O(\alpha_s v^2)$ order is also small.
we also give the relation of the cross sections and the $\mu_r$ in the
FIG.\ref{figure1}. There are about 10 percents differences in the total cross sections between $m_c=1.5$ and $m_c=1.4$. 
We find the uncertainty of the total cross sections from $m_c$ is not small. If we choose $\mu_r=2m_c$, we could give 
the relations of the ratios of different parts to total cross sections
with the $\sqrt{s}$ in the FIG.\ref{figure2}. We will find that the contributions from $\mathcal O(\alpha_s)$ and $\mathcal O(\alpha_s v^2)$ become important and the one from LO becomes small when  $\sqrt{s}$ is large, but there are just about 10 percents contribution in  $\mathcal O(\alpha_s v^2)$. The contribution from the $\mathcal O(\alpha_s v^2)$ order is small once again.
\begin{table}
\begin{tabular}
{|>{\centering}p{0.15\textwidth}|>{\centering}p{0.20\textwidth}|>{\centering}p{0.10\textwidth}|>{\centering}p{0.10\textwidth}
|>{\centering}p{0.10\textwidth}|>{\centering}p{0.10\textwidth}|>{\hfill}p{0.10\textwidth}<{\hfill\hfill}|}
\hlinew{1pt}
 m&  $\alpha_s(\mu_r)$  &  $\sigma_{LO}$  & $\sigma_{NLO(\alpha_s)}$ & $\sigma_{NLO(v^2)}$& $\sigma_{NLO(\alpha v^2)}$ & $\sigma$\\
\hline
 1.5&  $\alpha_s(\frac{\sqrt{s}}{2}) = 0.211$  &  $5.973$  & $6.645$  & $1.335$ & $0.416$&$14.369 $\\
\hline
 1.5 & $\alpha_s(2 m) = 0.259$  &  $9.000$  & $8.771$  & $2.011$ & $-0.017$& $19.726$\\
\hline
 1.4 & $\alpha_s(\frac{\sqrt{s}}{2}) = 0.211$  &  $6.526$  & $7.754$  & $1.591$ & $0.667$ &$16.538$\\
\hline
 1.4 & $\alpha_s(2 m_c) = 0.267$  &  $10.450$  & $10.989$  & $2.548$ & $0.1989$&$24.185$\\
\hlinew{1pt}
\end{tabular}
\caption{In the follow parameters: $\alpha(\sqrt{s}) = 1/137$, $\langle{\mathcal{O}_1}\rangle_{J/\psi}=1.719 {\rm GeV}^3$, $\langle {\mathcal{O}_1}\rangle_{\eta_c}=0.432, {\rm GeV}^3 $,$\langle v^2\rangle_{J/\psi} = 0.090$ , $\langle v^2\rangle_{\eta_c}= 0.119$,, We give the cross sections with different m and renormalization scale $\mu$.}
\label{tab:2}
\end{table}

\section{summary}
\label{summary}
In this work we have calculated the $\mathcal O(\alpha_s v^2)$ correction in detail for
the processes $e^+e^-\to J/\psi + \eta_c$ within the frame of NRQCD.
The result at $\mathcal O(\alpha_s v^2)$ order give about 6 percent contribution to the total theoretical prediction while
the $\mathcal O(\alpha_s)$ correction and $\mathcal O(v^2)$ are about 40 percent and 14 percent contribution respectively.
It indicates that the convergence in the double perturbative expansions in  QCD $\alpha_s$ and relativistic $v^2$
are very well for the theoretical calculation on the production rate of the process  $e^+e^-\to J/\psi+\eta_c$.
Up to $O(\alpha_s v^2)$ order, the theoretical prediction with quite large uncertainty from charm quark mass and renormalization scale
can describe the experimental measurement.

This work is supported by the National Natural Science Foundation of
China (Nos.~10979056 and 10935012), in part by DFG and NSFC (CRC 110) and CAS under Project No. INFO-115-B01.

\bibliography{paper}
\end{document}